# Original Paper

# The Trust in AI-Generated Health Advice (TAIGHA) Scale and Short Version (TAIGHA-S): Development and Validation Study

Marvin Kopka[1,2]*, Azeem Majeed[3], Gabriella Spinelli[4], Austen El-Osta[2,5], Markus Feufel[1]

[1] Division of Ergonomics, Department of Psychology and Ergonomics (IPA), Technische Universität Berlin, Berlin, Germany
[2] Self-Care Academic Research Unit (SCARU), School of Public Health, Imperial College London, London, United Kingdom
[3] Department of Public Health and Primary Care, Imperial College London, London, United Kingdom
[4] College of Engineering, Design and Physical Sciences, Brunel University of London, Uxbridge, United Kingdom
[5] School of Life Course and Population Sciences, King's College London, London, United Kingdom

Austen El-Osta and Markus Feufel share the last authorship.

# Abstract

Artificial Intelligence tools such as large language models are increasingly used by the public to obtain health information and guidance. In health-related contexts, following or rejecting AI-generated advice can have direct clinical implications. Existing instruments like the Trust in Automated Systems Survey assess trustworthiness of generic technology, and no validated instrument measures users' trust in AI-generated health advice specifically. This study developed and validated the Trust in AI-Generated Health Advice (TAIGHA) scale and its four-item short form (TAIGHA-S) as theory-based instruments measuring trust and distrust, each with cognitive and affective components. The items were developed using a generative AI approach, followed by content validation with 10 domain experts, face validation with 30 lay participants, and psychometric validation with 385 UK participants who received AI-generated advice in a symptom-assessment scenario. After automated item reduction, 28 items were retained and reduced to 10 based on expert ratings. TAIGHA showed excellent content validity (S-CVI/Ave=0.99) and CFA confirmed a two-factor model with excellent fit (CFI=0.98, TLI=0.98, RMSEA=0.07, SRMR=0.03). Internal consistency was high (α=0.95).

Convergent validity was supported by correlations with the Trust in Automated Systems Survey (r=0.67/−0.66) and users' reliance on the AI's advice (r=0.37 for trust), while divergent validity was supported by low correlations with reading flow and mental load (all |r|<0.25). TAIGHA-S correlated highly with the full scale (r=0.96) and showed good reliability (α=0.88). TAIGHA and TAIGHA-S are validated instruments for assessing user trust and distrust in AI-generated health advice. Reporting trust and distrust separately permits a more complete evaluation of AI interventions, and the short scale is well-suited for time-constrained settings.

# Introduction

Given the growing popularity, availability and performance of generative Artificial Intelligence (AI) tools such as Large Language Models (LLMs), the public are increasingly using these technologies to obtain health information and guidance for a variety of health-related tasks and decisions [1]. This growing reliance on AI-generated information is particularly consequential in health-related contexts, where following or rejecting an AI tool's advice can have personal, clinical and safety consequences, as well as broader impacts on healthcare systems [2–4]. The recent case of a ChatGPT user who was hospitalised for bromism after following advice on how to reduce salt intake demonstrated these risks [5]. Similarly, LLMs may also inadvertently spread misinformation when generating inaccurate or fabricated content [6,7]. Whereas such incidents exemplify potential dangers, the same technology also promises to make healthcare more efficient.

Emerging empirical evidence suggests that these risks are amplified by users' high levels of trust in AI-generated medical advice. A recent MIT study [8] found that patients often trust medical recommendations produced by AI systems more than those provided by human clinicians, even when the AI advice is demonstrably incorrect. Notably, participants were less likely to critically challenge AI-generated guidance and more inclined to follow it with confidence, raising concerns about overreliance and reduced skepticism in decision-making. This tendency is particularly problematic in health contexts, where misplaced trust may lead to harmful self-management behaviours, delayed clinical intervention, or inappropriate treatment decisions.

At a system level, for instance, LLMs may support patient empowerment, their decision-making, and health education in community settings [9–12]. For non-experts, determining the accuracy of information or advice provided by an AI decision support tool (DST) is often challenging, yet they must still decide whether to trust and then follow the AI-generated advice. Although trust in AI has been shown to be an important predictor of whether people will use AI tools and follow, or not, their advice [10,13], trust formation is context-dependent and may differ depending on personal and situational variables [14]. Measuring use-case specific trust is therefore important to understand how to best support users in making safe and informed decisions in each context. Because no existing instrument specifically measures the trust that lay users place in AI-generated health advice, we sought to develop and validate the Trust in AI-Generated Health Advice (TAIGHA) scale.

Research on the measurement of trust in technology has its roots in organizational trust measurement. In one of the earliest articles, Mayer and colleagues defined trust as "*the willingness to be vulnerable to the actions of another party based on the expectation that the other will perform a particular action important to the trustor*" [15]. According to this definition, trust comprises three dimensions: (i) competence, (ii) integrity, and (iii) benevolence [15]. An alternative conceptualisation agrees with the general definition but distinguishes between cognitive and affective trust [16]. Whereas cognitive trust refers to positive beliefs about the trustee's ability and reliability, affective trust refers to the emotional dimension and perceived care the trustee demonstrates toward the trustor [16].

Trust in technology is most commonly assessed using the Trust in Automated Systems Survey, which is the most frequently cited instrument in this field [17,18]. This scale builds on Mayer and colleagues' definition [15] and asks users to rate their perceptions of a technological system. This scale has several limitations: it tends to produce overly positive responses [19], comprises two subscales rather than a single construct as originally proposed [20], and lacks specificity due to its broad and context-independent wording [18]. Although other psychometrically validated instruments have since been developed, including the Human-Computer Trust Scale [21], the Semantic Differential Scale for AI Trust [22], and the Trust Scale for Explainable AI [23], these instruments share similar limitations. Most importantly, they are not sensitive to specific use cases as they measure general trust in technology rather than context-dependent trust.

As the reliance on AI-based DSTs continues to grow across diverse domains, the context in which they are used also impacts how trust should be measured. That is, a scale for evaluating trust in autonomous vehicles should include different questions than a scale on trust in an AI system that provides medical advice. In other words, the context of implementation is important to develop trust measurement scales with higher (use-case specific) applicability. Similarly, Schlicker et al. argue that most existing scales do not actually measure trust, but rather a general perceived trustworthiness [24]. Trustworthiness refers to properties of a system (e.g., the developer, the algorithms, its perceived accuracy), whereas trust refers to the attitude a person has towards this system. Most validated instruments, however, confound trustworthiness and trust and validated instruments for actual trust are missing, which is particularly relevant in health DST use cases, where users may perceive an AI tool as not trustworthy but still decide to follows its advice [24,25]. It is therefore worthwhile to consider how trust should be measured in AI systems deployed to provide healthcare advice to a lay audience. In such contexts, users typically have no visibility of the developer, the provenance of the data, or the type of algorithm employed, and their only relationship is with the technology itself. As a result, trust is not mediated through institutional or professional actors but is formed directly through interaction with the AI system. Thus, measurement approaches that capture behavioural reliance rather than perceived system qualities alone are needed.

In summary, although several instruments exist to measure trust in technology, most measure perceived trustworthiness with limited applicability to the context of AI-generated health advice. Trust in this domain forms against the background of a particularly high degree of uncertainty due to use-case specific factors related to personal safety risks, the varying accuracy and quality of medical evidence and the differing transparency of AI advice. Hence, due to the nature of the AI technology and the opacity of the underlying decision-making process, asking its users rather general questions related to whether "the system behaves in an underhanded manner" [17], may not be meaningfully answered in the context of AI health advice. To address this gap, we developed and validated the Trust in AI-Generated Health Advice (TAIGHA) scale.

To specifically measure trust rather than perceived trustworthiness [24], the TAIGHA scale builds on McAllister's definition and conceptualises trust as consisting of both cognitive and affective components [16]. We did not build on the definition by Mayer et al., because it mostly contains attributes related to trustworthiness rather than trust [15]. The scale further builds on results from newer psychometric analyses of the original Trust in Automated Systems Survey, which identified two orthogonal and non-diametrically opposed dimensions: trust and distrust [20,26]. Accordingly, the TAIGHA scale provides a use-case-specific measure of trust in AI that includes two subscales, trust and distrust, each of which comprises cognitive and affective items. In addition to the main scale, we aimed to develop a short version, the TAIGHA-S, to measure trust and distrust in time-sensitive scenarios using as few items as possible. In applied research, trust is often measured as a secondary outcome using self-developed, single-item measures [27,28]. The TAIGHA-S is thus intended to offer researchers and practitioners a psychometrically validated alternative for situations in which they want to assess trust and distrust in AI-generated health advice and would otherwise rely on unvalidated questions.

The primary aim of this study was to develop and validate the TAIGHA scale, a psychometrically sound, context-specific instrument measuring trust and distrust in AI-generated health advice. Further to evaluating the factor structure, reliability and validity of the TAIGHA scale, we sought to develop and validate a short version (TAIGHA-S) for practical or large-scale use.

# Methods

**Ethics**

This study received ethical approval from the Ethics Committee of the Department of Psychology & Ergonomics at Technische Universität Berlin (#2708652). The study and procedures adhered to the Declaration of Helsinki.

**Instrument Development**

To develop the TAIGHA items, we used a generative AI-based method called AI-GENIE to develop an initial set of items and reduce this item set. In a manual validation, we then tested the items, reduced the item set further, and assessed validity and reliability. The AI-GENIE generative AI-based method [29], which uses an LLM to generate several candidate items and then obtain vector embeddings (numerical representations of text) for these items using an encoder, was used to develop the baseline version of the TAIGHA scale. More information on this approach can be found in [29]. In this process, the AI-GENIE represents the contextual meaning of each item numerically, which allows it to cluster items with similar meanings. For the vector embeddings, we used OpenAI's text-embedding-3-small. Next, the psychometric properties of the scale were assessed by AI-GENIE using a network-based method called exploratory graph analysis, which analyzes the vector embeddings to identify a factorial structure [30]. Subsequently, the normalized mutual information (NMI) value – which indicates the proportion of items classified into the same factor as in the input factorial structure [31] – was automatically calculated. At this stage, the NMI is expected to be relatively low, as no item selection has yet been applied. Additionally, the item pool may contain items that are phrased very similarly. To address this, unique variable selection was applied by AI-GENIE to remove items that were too similar in wording [32]. In the proceeding step, bootstrapped exploratory graph analysis was used by AI-GENIE to select only those items that were stable, i.e., items that consistently get grouped into the same factor. Here, item stability is calculated as a metric to determine how often an item gets grouped into the same category. After the item pool has been reduced using bootstrapped exploratory graph analysis, the reduced item pool was validated again using NMI and item stability by AI-GENIE.

We used GPT-4.1 within AI-GENIE to generate the items, because this model follows instructions consistently [33]. We applied the trust theory consisting of trust and distrust as two orthogonal factors [20,26]. For the LLM input, trust was defined as the "willingness to follow health advice from an AI because you expect it to help you", and included cognitive items measuring the "expectation that following the AI's advice will lead to positive health outcomes", and affective items measuring "positive feelings such as comfort, reassurance, or peace of mind when considering or relying on the AI's advice itself". Distrust was defined as a "protective stance, avoiding this piece of health advice because you expect it to harm you", and included cognitive items measuring the "expectation that following the AI's advice will lead to negative health outcomes", and affective items measuring "negative feelings such as anxiety, unease, or worry when considering or relying on the AI's advice".

Although the automated validation procedure with the AI-GENIE may yield results that are highly similar to empirical validations [29] and on average produces even better items than those written by humans [34], empirical validation is still essential to generate valid scales. For example, the AI-GENIE approach can establish a factorial structure but generates items based solely on the input theory. Hence, the construct and content validity of the resulting items must be verified empirically. To include only items with high content validity (as judged by experts) and acceptable psychometric properties, we therefore conducted a traditional psychometric evaluation in the next step.

**Validation Procedure**
We followed the validation procedure by Yusoff et al. [35]: first, we conducted content validation by surveying domain experts to determine whether each item was relevant and representative of the trust and distrust constructs. Second, we performed face validation by asking intended end users (i.e., general survey participants without psychometric or domain expertise) to evaluate whether each item was clear and comprehensible. Finally, after selecting the final item set, we conducted a psychometric evaluation where participants received health advice from an AI, made a health-related decision, and completed the TAIGHA scale along with related and unrelated questionnaires. Using these responses, we determined construct validity through correlations with related and unrelated measures, conducted a confirmatory factor analysis (CFA) to assess whether the factorial structure fits with our theoretical structure, and also assessed reliability using Cronbach's Alpha and McDonald's Omega.

In the first step, ten domain experts completed an online questionnaire hosted on SoSciSurvey. They received an introduction to the theory the TAIGHA scale is based on, reviewed the items and rated each item's relevance and representativeness on a 4-point Likert scale.

In the second step, 30 non-experts completed a similar online questionnaire. They received a brief explanation of the study's purpose, a short introduction to the theory and an example of AI-generated health advice. They were then asked to rate each item's clarity and comprehensibility on a 4-point Likert scale.

In the third step, we used a symptom-assessment scenario previously used in experimental and observational studies [9]. We specifically selected this scenario because it represents a real-world use case of obtaining health advice from an AI-based DST, which is common among medical lay users [36–39]. We used a set of validated case scenarios that describe real-world patient cases in which medical lay users sought online health information to assist them in deciding whether and where to seek care [40,41]. After providing informed consent and demographic information, participants received one randomly selected scenario from the full set of 27 scenarios and were asked to indicate whether they would (i) seek emergency care for the described symptoms (i.e., call 999 or visit the emergency room), (ii) seek non-emergency care (i.e., trying to see a doctor within the next days), or (iii) engage in self-care (i.e., let the health issue get better on its own and review the situation in a few days again). They then received AI-generated health advice on what to do in this scenario (generated using GPT o3) and were asked to choose again among the three options. Finally, participants were asked to complete the TAIGHA scale, as well as several additional questionnaires measuring related and unrelated constructs.

## Participants
Experts for the content validation were recruited via snowball sampling from the researchers' professional network. Following Yusoff et al. [42], we aimed to include ten experts, who participated between 27 May 2025 and 11 June 2025.

Participants for the face validation were recruited via Prolific [43] using a random sample of English-speaking users from the UK. In line with Yusoff et al. [44], we aimed to include 30 participants. Data for the face validation were collected on 11 June 2025.

Participants for the psychometric validation were also recruited via Prolific using a random sample of English-speaking users from the UK. Since there are no clear guidelines for sample size in CFA, but at least 300 participants are generally recommended [45], the study oversampled by 100 participants to account for the exclusion of participants that did not answer two embedded attention checks correctly. The recruitment target was a total of 400 participants by 29 September 2025.

## Instrument Validation
All analyses were performed in R 4.3.3, using the tidyverse packages [46] and the psych package [47].

### Content & Face Validity
To determine content validity, we used the Content Validity Index (CVI) [42]. To calculate this index, participants' responses on the relevancy and representativeness of the items were dichotomised: *not* or *somewhat relevant* were coded as 0, whereas *quite* or *highly relevant* were coded as 1. The CVI was calculated for each item separately, and items had to reach a standard cutoff value of at least 0.80 to remain part of the scale [42]. Additionally, we calculated scale-wide average CVI (S-CVI/Ave), and the scale-wide CVI with universal agreement (S-CVI/UA). The S-CVI/Ave is a metric to determine the average CVI of the whole scale, whereas the S-CVI/UA indicates the percentage of items rated as relevant by all experts. Both indices were required to reach a standard minimum cutoff value of 0.80 [42]. If these cutoffs were not met, items with the lowest CVI values were iteratively removed until acceptable values were reached.

To determine face validity, we used the face validity index (FVI) [44]. This index is calculated analogously to the CVI based on end-users' ratings of item clarity and comprehensibility. We determined the FVI, the scale-wide average FVI (S-FVI/Ave), and the scale-wide FVI with universal agreement (S-FVI/UA). A cutoff of 0.80 or higher was considered acceptable for all validity metrics [44].

### Construct Validity

To determine construct validity, we measured the correlations of the TAIGHA scale with related questionnaires (to determine convergent validity), and with unrelated questionnaires (divergent validity). For convergent validity, we used the most widely used Trust in Automated Systems Survey [17], the Propensity to Trust in Technology Scale [48]. Because the AI may have given the same recommendation as participants' initial appraisal, we operationalized reliance as participants' shifts, i.e., included cases in which the AI's advice and participants' initial appraisal differed and determined whether or not they had changed their own appraisal in favor of the AI advice [49]. Correlations above 0.30 were considered moderate and thus evidence of a related construct, whereas values above 0.90 were considered too high, as they would measure nearly the same construct and a new questionnaire would not be needed [50].

To determine divergent validity, we used the Reading Flow Short Scale [51], the General Self-Efficacy Short Scale [52], and the NASA Task Load Index (NASA-TLX) [53]. We specifically chose these questionnaires as they may be used to assess the use of DST, but there is no evidence that trust in AI-generated health advice should be correlated with reading flow, self-efficacy or task load. However, we excluded the self-rated performance and frustration subscales of the NASA-TLX, as these were expected to be at least somewhat related to trust and reliance on AI advice. Correlations below 0.30 were considered low and thus evidence of high divergent validity [50].

### Criterion Validity

To determine criterion validity, we measured the correlations of the TAIGHA scale with the outcome it aims to predict, that is, participants' reliance on the AI's health advice (also referred to as trust behavior or advice-taking). Again, correlations above 0.30 were considered moderate and thus evidence of criterion validity [50].

### Confirmatory Factor Analysis

We conducted a CFA using maximum likelihood estimation to assess whether the items show the same factorial structure as proposed by the underlying theory. The model fit was evaluated using commonly applied goodness-of-fit indices [35,54]: as incremental fit indices, we used the goodness of fit index (GFI), the comparative fit index (CFI), the Tucker-Lewis Index (TLI), and the Normed Fit Index (NFI). As absolute fit indices, we used the root mean square error of approximation (RMSEA), and the standardized root mean squared residual (SRMR).

### Reliability

Lastly, we determined the reliability of the TAIGHA scale using measures of internal consistency, i.e., the degree to which items within each subscale yield similar responses [55]. We calculated Cronbach's Alpha [56] and McDonald's Omega [57] to quantify internal consistency. Values above 0.75 were considered acceptable [58].

### Development of Short Scale

To develop the short version (TAIGHA-S), we followed the approach introduced in a previous article [59], which was originally based on the development of a short scale by Wessel et al. [60]. To make sure that the scale can be used without high time demands, we aimed to include two items per subscale [61]. Based on our underlying theory, we aimed to include one cognitive (dis-)trust and one affective (dis-)trust item each. To select items, we first used the item-total correlation of each item (measuring the correlation between each item and the total score when that item was excluded [62]) and selected items with the highest values. Second, when multiple items had identical item-total correlations, we used factor loadings from the CFA to select the candidate items that should be included. To validate the short scale, we correlated it with the full version and repeated the full validation process: assessing convergent and divergent validity, conducting a CFA and determining reliability.

# Results

### Sample Characteristics

For the content validation, we surveyed 10 domain experts: eight (80%) were postdoctoral researchers, one (10%) an assistant professor, and one (10%) an associate or full professor. Three experts (30%) reported their primary field of expertise to be medicine and healthcare, three (30%) human-computer interaction, two (20%) psychology, and two (20%) AI or computer science. Four (40%) reported having conducted research on trust, six (60%) on human-computer interaction, seven (70%) on digital health, and seven (70%) on questionnaire development. For the face validation, we surveyed 30 participants who were exactly balanced in gender and on average 41 years old (SD=14). For the psychometric evaluation, we surveyed 393 participants, of whom eight (2%) were excluded for answering attention check questions incorrectly. The characteristics of the final included sample are shown in **Table 1**.

**Table 1. Characteristics of the Sample Included for the Psychometric Evaluation.**

| Characteristic | Total |
|---|---|
| **Age (years), M(SD)** | 30 (16) |
| **Gender, n (%)** | |
|   Male | 184 (47.8%) |
|   Female | 199 (51.7%) |
|   Other | 1 (0.0%) |
|   Prefer not to say | 1 (0.0%) |
| **Education, n (%)** | |
|   Finished high school with no qualifications | 5 (1.3%) |
|   Secondary school-leaving certificate | 14 (3.6%) |
|   High school diploma | 50 (13.0%) |
|   Completed apprenticeship | 13 (3.4%) |
|   Vocational secondary diploma | 32 (8.3%) |
|   A-Levels | 38 (9.9%) |
|   University degree | 233 (60.5%) |
| **Self-Efficacy, M(SD)[a]** | 4.0 (0.8) |
| **Trust in Technology, M(SD)[a]** | 3.6 (0.5) |

[a] On a 5-point Likert scale

## Item Generation with AI-GENIE

Initially, 60 items were generated. The overall questionnaire had an NMI of 90.6% (60.0% for the trust subscale and 67.9% for the distrust subscale) and item stability values between 5% and 100%. This indicates poor psychometric properties. After the automated item selection, 28 items were left and the reduced questionnaire showed good psychometric properties with an NMI value of 100% (100% for both subscales) and item stability values between 81% and 100%. The automatically reduced questionnaire was then given to domain experts to further reduce the items to include only items that demonstrate high content validity.

## Content Validity

The CVI-I values of the reduced questionnaire ranged from 0.7 to 1.0, see Table 2. Because 8 items did not reach the cutoff of 0.8, those items were deleted. The S-CVI/Ave was 0.90 and therefore acceptable, but the S-CVI/UA was 0.45 and not acceptable. Following a conservative approach, items with a CVI-I of 0.8 were then deleted to improve content validity. This resulted in acceptable values of S-CVI/Ave=0.99 and S-CVI/UA=0.90.

**Table 2. Content Validity Indices of the Initial 28 Items.**

| Item | Rated as not relevant (n) | Rated as relevant (n) | CVI-I |
|---|---|---|---|
| Initial Trust Item 1 | 3 | 7 | 0.70 |
| Initial Trust Item 2 | 0 | 10 | 1.00 |
| Initial Trust Item 3 | 0 | 10 | 1.00 |
| Initial Trust Item 4 | 3 | 7 | 0.70 |
| Initial Trust Item 5 | 2 | 8 | 0.80 |
| Initial Trust Item 6 | 2 | 8 | 0.80 |
| Initial Trust Item 7 | 0 | 10 | 1.00 |
| Initial Trust Item 8 | 2 | 8 | 0.80 |
| Initial Trust Item 9 | 3 | 7 | 0.70 |
| Initial Trust Item 10 | 3 | 7 | 0.70 |
| Initial Trust Item 11 | 2 | 8 | 0.80 |
| Initial Trust Item 12 | 4 | 6 | 0.60 |
| Initial Trust Item 13 | 0 | 10 | 1.00 |
| Initial Trust Item 14 | 3 | 7 | 0.70 |
| Initial Trust Item 15 | 1 | 9 | 0.90 |
| Initial Trust Item 16 | 2 | 8 | 0.80 |
| Initial Trust Item 17 | 2 | 8 | 0.80 |
| Initial Distrust Item 1 | 2 | 8 | 0.80 |
| Initial Distrust Item 2 | 0 | 10 | 1.00 |
| Initial Distrust Item 3 | 2 | 8 | 0.80 |
| Initial Distrust Item 4 | 0 | 10 | 1.00 |
| Initial Distrust Item 5 | 2 | 8 | 0.80 |
| Initial Distrust Item 6 | 0 | 10 | 1.00 |
| Initial Distrust Item 7 | 2 | 8 | 0.80 |
| Initial Distrust Item 8 | 0 | 10 | 1.00 |
| Initial Distrust Item 9 | 3 | 7 | 0.70 |
| Initial Distrust Item 10 | 3 | 7 | 0.70 |
| Initial Distrust Item 11 | 0 | 10 | 0.10 |

**Face Validity**

All FVI-I values of the included items were acceptable, ranging from 0.7 to 1.0, see **Table 3**. With S-CVI/Ave=0.99 and S-CVI/AU=0.80, scale-wide face validity was acceptable as well.

**Table 3. Face Validity Indices of the Items remaining after Content Validation.**

| Item | Rated as not clear and understandable (n) | Rated as clear and understandable (n) | FVI-I |
|---|---|---|---|
| **Trust Items** | | | |
| 1: I feel comfortable about following the AI's health advice | 0 | 30 | 1.00 |
| 2: The AI provides advice that should lead to better health. | 1 | 29 | 0.97 |
| 3: I feel a sense of trust toward the AI's advice | 0 | 30 | 1.00 |
| 4: I feel positive about relying on the AI's advice. | 0 | 30 | 1.00 |
| 5: I believe the AI's advice will help me make good health decisions. | 0 | 30 | 1.00 |
| **Distrust Items** | | | |
| 1: I feel uncomfortable with the idea of following the AI's advice. | 0 | 30 | 1.00 |
| 2: I think following the AI's advice could pose health risks. | 0 | 30 | 1.00 |
| 3: I feel distressed about relying on the AI's advice. | 0 | 30 | 1.00 |
| 4: I feel tense about taking the AI's advice into account. | 0 | 30 | 1.00 |
| 5: The idea of trusting this AI advice makes me uneasy. | 2 | 28 | 0.93 |

## Construct Validity

All items had few (or no) missing responses, low skewness and kurtosis; **Table 4.** The item difficulty was within acceptable values, and all items had high item-total correlation.

**Table 4. Descriptive Statistics, Item Difficulty, and Item-Total Correlation of the Included Items.**

| Item | Missing data (%) | Mean (SD) | Median (IQR) | Skewness | Kurtosis | Min - Max | Item Difficulty | Item-Total Correlation |
|---|---|---|---|---|---|---|---|---|
| Trust Item 1 | 0% | 3.78 (0.90) | 4 (3-4) | -0.9 | 0.9 | 1-5 | 0.76 | 0.84 |
| Trust Item 2 | 0% | 3.77 (0.76) | 4 (3-4) | -0.6 | 0.6 | 1-5 | 0.75 | 0.78 |
| Trust Item 3 | 0.3% | 3.55 (0.89) | 4 (3-4) | -0.9 | 0.6 | 1-5 | 0.71 | 0.85 |
| Trust Item 4 | 0.3% | 3.58 (0.92) | 4 (3-4) | -0.7 | 0.2 | 1-5 | 0.72 | 0.85 |
| Trust Item 5 | 0% | 3.74 (0.81) | 4 (3-4) | -0.9 | 1.3 | 1-5 | 0.75 | 0.85 |
| Distrust Item 1 | 0% | 2.32 (1.06) | 2 (2-3) | 0.8 | -0.2 | 1-5 | 0.47 | 0.80 |
| Distrust Item 2 | 0% | 2.14 (0.95) | 2 (2-3) | 0.9 | 0.4 | 1-5 | 0.43 | 0.79 |
| Distrust Item 3 | 0% | 1.96 (0.94) | 2 (1-2) | 1.1 | 1.2 | 1-5 | 0.39 | 0.79 |
| Distrust Item 4 | 0.3% | 2.16 (1.05) | 2 (1-3) | 0.9 | 0.1 | 1-5 | 0.43 | 0.86 |
| Distrust Item 5 | 0% | 2.29 (1.07) | 2 (2-3) | 0.8 | 0.0 | 1-5 | 0.46 | 0.85 |

### *Convergent Validity*

The trust and distrust subscales showed a very high correlation (r=0.67 [95%CI 0.61 to 0.72] and r=-0.66 [95%CI -0.60 to -0.71], respectively) with the Trust in Automated Systems Survey, as well as high correlations with the Propensity to Trust Scale (r=0.54 [95%CI 0.46 to 0.61] and r=-0.46 [95%CI -0.38 to -0.53]), see **Table 5**.

**Table 5. Convergent Validity. Correlations with other variables that measure related concepts.**

|  | Trust Items | Distrust Items | Trust in Automated Systems Survey | Propensity to Trust Scale |
|---|---|---|---|---|
| Trust Items | 1 |  |  |  |
| Distrust Items | 0.78 | 1 |  |  |
| Trust in Automated Systems Survey | 0.67 | -0.66 | 1 |  |
| Propensity to Trust Scale | 0.54 | -0.46 | 0.43 | 1 |

### *Divergent Validity*

The trust and distrust subscales showed low to very low correlations with unrelated items. The highest correlation was between the NASA-TLX mental demand subscale and the distrust items (r =0.25 [95%CI 0.15 to 0.34]), and the lowest correlation was between the trust subscale and the NASA-TLX effort subscale (r=-0.01 [95%CI -0.11 to 0.09]); see **Table 6**.

**Table 6. Divergent Validity. Correlations with other variables that measure unrelated concepts.**

|  | Trust Items | Distrust Items | Reading Flow Short Scale | General Self-Efficacy Short Scale | (NASA-TLX) | | | |
|---|---|---|---|---|---|---|---|---|
|  |  |  |  |  | Mental Demand | Physical Demand | Temporal Demand | Effort |
| Trust Items | 1 |  |  |  |  |  |  |  |
| Distrust Items | -0.78 | 1 |  |  |  |  |  |  |
| Reading Flow Short Scale | 0.19 | -0.19 | 1 |  |  |  |  |  |
| General Self-Efficacy Short Scale | 0.14 | -0.12 | 0.22 | 1 |  |  |  |  |
| Mental Demand (NASA-TLX) | -0.13 | 0.25 | -0.03 | -0.09 | 1 |  |  |  |
| Physical Demand (NASA-TLX) | -0.10 | 0.21 | -0.16 | -0.15 | 0.49 | 1 |  |  |
| Temporal Demand (NASA-TLX) | -0.09 | 0.17 | -0.22 | -0.20 | 0.46 | 0.57 | 1 |  |
| Effort (NASA-TLX) | -0.01 | 0.07 | 0.08 | -0.04 | 0.43 | 0.23 | 0.20 | 1 |

### *Criterion Validity*

The trust subscale showed a moderate correlation with reliance on AI advice (r = 0.37 [95%CI 0.22 to 0.50]), whereas the distrust subscale showed a low correlation with reliance on AI advice (r = -0.19 [95%CI -0.03 to -0.34]). Correlations of the trust subscale with reliance on AI advice were higher than the correlations of the Trust in Automated Systems Survey (0.26 [95%CI 0.10 to 0.40]) and the Propensity to Trust Scale (r = 0.09 [95%CI -0.07 zo 0.24]) with reliance on AI advice; see **Table 7**.

**Table 7. Criterion Validity. Correlations with the outcome it aims to predict.**

|  | Reliance on AI advice |
|---|---|
| Trust Items | 0.37 |
| Distrust Items | -0.19 |
| Trust in Automated Systems Survey | 0.26 |
| Propensity to Trust Scale | 0.09 |

**CFA**

The proposed two-factor model fit the data well, see **Table 8** for goodness-of-fit metrics. The standardised factor loadings ranged from 0.80 to 0.90, and, as expected, both factors had a high correlation but were distinct factors (r=-0.84), see **Figure 1**.

**Table 8. Fit Indices for the proposed two-factor model.**

| Fit index | Cutoff | Value |
|-----------|--------|-------|
| GFI | > 0.90 | 0.95 |
| CFI | > 0.90 | 0.98 |
| TLI | > 0.90 | 0.98 |
| NFI | > 0.90 | 0.97 |
| RMSEA | < 0.08 | 0.07 |
| SRMR | < 0.08 | 0.03 |

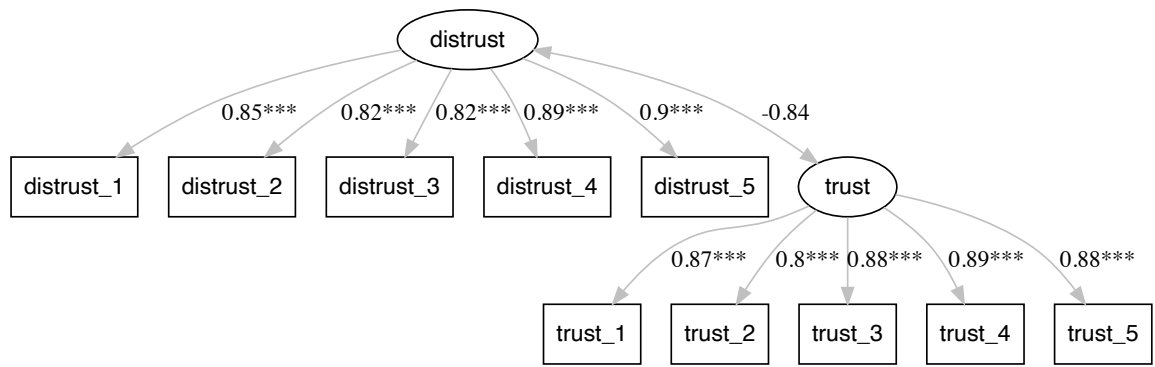

**Figure 1. Factor Structure and Standardized Factor Loadings.**

**Reliability**

The internal consistency of the TAIGHA scale was excellent, with Cronbach's Alpha=0.95 and McDonald's Omega=0.96. The two subscales demonstrated high internal consistency as well (α=0.94 and ω=0.94 for the trust subscale, and α=0.93 and ω=0.93 for the distrust subscale).

**Short Scale**

For the trust subscale, item five had the highest ITC among cognitive trust items, and items three and four had the highest ITC among affective trust items (ITC=0.85 for all three items). Item four had a slightly higher factor loading (0.89) than item three (0.88) and was therefore chosen for the short scale. For the distrust subscale, item two had the highest ITC among cognitive distrust items (ITC=0.79), and item four had the highest ITC among affective distrust items (ITC=0.86). The TAIGHA-S short scale thus consists of items four and five for the trust subscale and items two and four for the distrust subscale.

The short scale and its subscales showed a high correlation with the full scale and subscales (**Table 9**). It also demonstrated high convergent validity through high correlations with related concepts, high criterion validity through a moderate correlation with reliance on AI advice (**Table 9**) and high divergent validity through low correlations with unrelated concepts (**Table 10**).

**Table 9. Convergent and Criterion Validity. Correlations with other variables that measure related concepts and with the outcome it aims to predict.**

| Unrelated Construct | Trust Short Subscale | Distrust Short Subscale |
|---|---|---|
| Full Trust Items | 0.96 | 0.96 |
| Full Distrust Items | -0.75 | -0.73 |
| Trust in Automated Systems Survey | 0.63 | -0.61 |
| Propensity to Trust Scale | 0.52 | -0.40 |
| Reliance on AI Advice | 0.36 | -0.14 |

**Table 10. Divergent Validity. Correlations with other variables that measure unrelated concepts.**

| Unrelated Construct | Trust Short Subscale | Distrust Short Subscale |
|---|---|---|
| Reading Flow Short Scale | 0.17 | -0.18 |
| General Self-Efficacy Short Scale | 0.12 | -0.10 |
| Mental Demand (NASA-TLX) | -0.13 | 0.25 |
| Physical Demand (NASA-TLX) | -0.09 | 0.22 |
| Temporal Demand (NASA-TLX) | -0.05 | 0.18 |
| Effort (NASA-TLX) | -0.03 | 0.07 |

Similar to the full scale, the CFA demonstrated a good fit, with factor loadings ranging between 0.84 and 0.90 (**Figure 2**), and all goodness-of-fit metrics reaching cutoff values (**Table 11**).

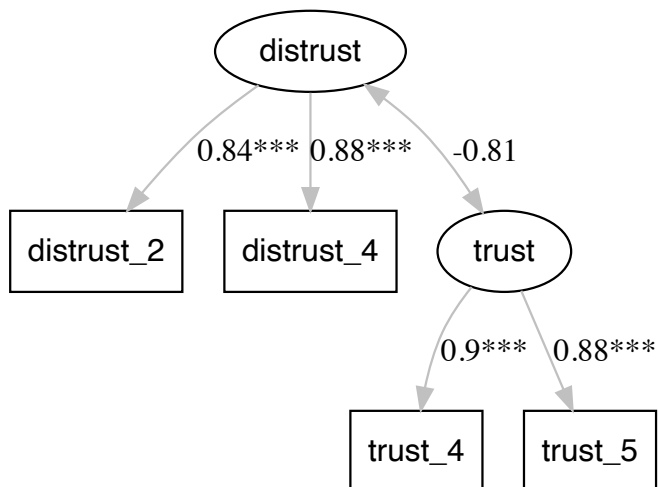

**Figure 2. Factor Structure and Standardized Factor Loadings of the Short Scale.**

**Table 11. Fit Indices for the short scale.**

| Fit index | Cutoff | Value |
| --- | --- | --- |
| GFI | > 0.90 | 0.99 |
| CFI | > 0.90 | 1.00 |
| TLI | > 0.90 | 1.00 |
| NFI | > 0.90 | 0.99 |
| RMSEA | < 0.08 | 0.00 |
| SRMR | < 0.08 | 0.01 |

The reliability of the short scale was acceptable ($\alpha$=0.88 and $\omega$=0.93 for the full scale, $\alpha$=0.88 and $\omega$=0.85 for the trust subscale, and $\alpha$=0.84 and $\omega$=0.89 for the distrust subscale).

# Discussion

**Principal Findings**

This study aimed to develop and validate a measurement instrument that specifically assesses trust in AI-generated health advice. Given the limitations of existing questionnaires that measure only general trust in technology, lack contextual sensitivity and assess perceived trustworthiness of systems rather than the user's willingness to follow a specific piece of health advice generated by an AI tool [17,24], we developed and validated the TAIGHA scale and its short form, TAIGHA-S.

Both scales were explicitly designed for AI-generated health advice and are based on empirical findings from previous trust scales and McAllister's trust theory, that is, they distinguish between trust and distrust [20,26], considering both cognitive and affective items [16]. The TAIGHA scale demonstrated high content validity as rated by experts from multiple relevant domains and was found to be face valid, that is, understandable to a lay audience. It also showed excellent reliability, very high goodness-of-fit indices for a two-factor model, and high convergent and divergent validity. Thus, the TAIGHA scale appears to measure users' trust in AI-generated health advice reliably without capturing unrelated constructs. Notably, in our criterion validation, the TAIGHA scale correlated more strongly with whether participants followed the AI's advice than the existing general-purpose Trust in Automated Systems Survey [17]. More broadly, this finding suggests that a scale developed for a specific health advice-taking context can explain behavior in that context more accurately than a general technology trust scale that lacks items adapted to the use case.

Our findings extend and combine several lines of research. First, the TAIGHA scale provides an example of how to operationalize trust rather than trustworthiness - two concepts that many prior studies have conflated [24]. Much of the literature on trust in technology measures perceived trustworthiness attributes of the system (such as reliability or competence of a technology) instead of the user's state of trusting and acting on the advice it provides [14,24]. To address this, we applied McAllister's theoretical trust model [16], which focuses on cognitive and affective components within the trustor (lay users, in this case), rather than Mayer et al.'s model, which emphasises trustworthiness and characteristics of the trustee such as competence, integrity, and benevolence [15,63]. This conceptual difference likely explains why the TAIGHA scale correlated more strongly with advice-taking behavior than the more general Trust in Automated Systems Survey, which is based on Mayer et al.'s model. Such a focus is especially relevant in health contexts, where AI may sometimes not be perceived as trustworthy, yet users still follow its advice [25].

Second, the TAIGHA scale treats distrust as an orthogonal dimension rather than the mere opposite of trust. Re-analyses of previous trust scales have repeatedly shown that separate trust and distrust factors better explain the factor structure of trust questionnaires than a single unidimensional construct [20,26]. Our results are consistent with this view and may suggest that interventions aimed at reducing distrust (e.g., by citing sources or providing explanations) may not necessarily increase trust. Conversely, interventions designed to increase trust may not automatically reduce users' skepticism or caution toward AI tools. Evaluating both subscales therefore enables researchers and developers to determine which component an intervention primarily affects or should be addressed.

Third, we employed a relatively new methodological approach that integrates generative AI into psychometrics [29,64–66]. Using the AI-GENIE framework, we first generated an item pool with GPT-4.1 and then selected items that demonstrated strong psychometric properties [29]. Prior research indicates that items generated by LLMs are often of higher quality than those written by humans [34], and that the automated validation procedure can produce items that also perform well in traditional empirical evaluations. Our results provide further evidence for the latter, as the automated validation process was technically successful. However, subsequent content validation by domain experts seems to remain important, as several psychometrically sound items in this study were removed for being irrelevant or not fitting to theory. This may be the result of LLMs' high output variability and their limited adherence to instructions [67–69]. Future research using the AI-GENIE approach should therefore iteratively refine prompts to ensure that the generated items fit well with the underlying theory and include a manual item selection phase to remove irrelevant items. Although such an automated approach can substantially accelerate item generation and validation, our results underscore that a manual construct and content validation remain important to guarantee the quality and meaningfulness of the scale items.

From a theoretical perspective, our findings extend the two-factor perspective (trust and distrust) from general trust theory [16] and general trust in technology research [17,20,26] to a domain-specific trust measurement. Second, they reinforce the argument that trust in AI advice among lay users should be behavior-proximal [24], that is, scales should ask users about their cognition and affect in the actual decision context rather than about the AI characteristics and their perceived trustworthiness, which was shown to improve correlation with reliance. Third, the affective components of trust, which have often been excluded from traditional technology trust scales [17,23], have been shown to be an important predictor of trusting behavior, which may provide new opportunities for trust interventions and AI design.

From a practical standpoint, the TAIGHA scale offers a highly relevant outcome measure for evaluating public-facing AI tools. When trust is included only as a secondary outcome, the short form, TAIGHA-S, can be used as an alternative to self-constructed single-item measures, as it has low resource demands but good psychometric properties [61]. Reporting trust and distrust separately also supports safety evaluations: in situations in which users rely excessively on low-quality or high-risk advice or rely insufficiently on high-quality advice, the reasons for that behavior can be identified more precisely. This could help inform interventions that address either trust or distrust to promote safer and more appropriate reliance on AI-generated health advice.

By capturing trust and distrust as separate constructs, it is possible to uncover the underlying reasons for user behaviour, identify potential safety risks, and design interventions that strengthen trust in reliable AI outputs and/or mitigate distrust that limits appropriate use. Furthermore, validated tools such as TAIGHA-S provide a low-resource yet psychometrically robust method for evaluating these perceptions, enabling efficient assessment of AI tools in healthcare and supporting their safe and effective adoption. This is particularly important in contexts where the growing use of large language model (LLM) technologies serves as a source of healthcare guidance for lay users, particularly in light of the persistent resource constraints faced by healthcare systems.

### Limitations
The development and validation of the TAIGHA scale have several limitations. Although we used a relatively high sample size of 385 participants [45,70], the principal limitation of this study is that the online-survey format may have resulted in a sample with greater technological affinity than the general UK population, as may be the case with research panels. Despite this, generalisability should not be substantially impaired since AI tools are deployed through technological systems, and their users are often more tech-savvy. However, the transferability of our findings to contexts in which new users engage with unfamiliar technology for the first time may be limited.

Another limitation arises from the mode of interaction. Because participants did not engage with the AI in a long-term scenario, we acknowledge that trust may also change over time [14]. Although this may constrain generalisability to long-term interactions, in many scenarios such as assessing acute symptoms, users typically interact only once with an AI DST rather than repeatedly [71]. Future studies should examine the applicability and validity of the TAIGHA scale, particularly its convergent validity and its association with reliance in longitudinal or repeated-use settings.

We acknowledge that using scenarios made the empirical validation somewhat artificial as these are not real-world interactions. However, this approach represents an appropriate trade-off between data quality and ethical feasibility, as asking participants experiencing real acute symptoms to use an AI tool and answer multiple questions would raise ethical concerns. To approximate real-world behavior, the use case scenarios and experimental setup in this study were validated in previous research, included real patient cases, and were shown to be representative with respect to the use case [40,41]. Nevertheless, it would be valuable to further validate the TAIGHA scale in future studies using different health advice scenarios, such as those involving self-care or health promotion. Also, the scale focused on assessing trust in AI by lay users. Future studies should explore to what extent this scale may also extend to assess trust in expert-facing AI tools.

Lastly, although the TAIGHA scale showed a moderate correlation with reliance and a stronger association with reliance than the Trust in Automated Systems Survey, its correlation with and thus predictive power for actual behavior remains limited. This outcome is expected, as decisions are also influenced by additional factors, such as individuals' heuristics when taking advice into account, their integration of the advice with pre-existing knowledge and experience, and situational circumstances [9,14,72,73]. Therefore, researchers should, whenever possible, include measures of actual behavior alongside trust assessments.

## Conclusions
In this article, we introduced and validated the TAIGHA scale and its four-item short form (TAIGHA-S) to assess users' trust and distrust in AI-generated health advice. Both instruments demonstrated excellent psychometric properties and correlated more strongly with advice-taking than a generic technology trust scale which indicates added value for health-advice contexts. The TAIGHA and TAIGHA-S can serve as outcome measures for evaluating public-facing, and potentially clinician-facing, AI systems. Reporting trust and distrust as separate constructs allows researchers to identify whether an intervention increases acceptance of AI advice, reduces aversion to it, or both. Overall, the TAIGHA scales provide a theory-grounded, use-case-specific measure that can help identify when people accept or reject AI-generated health advice.

**Conflicts of Interest**
None declared.

**Abbreviations**
AI: Artificial Intelligence
CFA: Confirmatory Factor Analysis
CFI: Comparative Fit Index
CVI: Content Validity Index
CVI-I: Item-Level Content Validity Index
DST: Decision Support Tool
FVI: Face Validity Index
FVI-I: Item-Level Face Validity Index
GFI: Goodness of Fit Index
IQR: Interquartile Range
ITC: Item-Total Correlation
LLM: Large Language Model
M (SD): Mean (Standard Deviation)
NASA-TLX: NASA Task Load Index
NFI: Normed Fit Index
NMI: Normalized Mutual Information
RMSEA: Root Mean Square Error of Approximation
SRMR: Standardized Root Mean Squared Residual
S-CVI/Ave: Scale-Level CVI, Average
S-CVI/UA: Scale-Level CVI, Universal Agreement
S-FVI/Ave: Scale-Level FVI, Average
S-FVI/UA: Scale-Level FVI, Universal Agreement
TAIGHA: Trust in AI-Generated Health Advice (Scale)
TAIGHA-S: TAIGHA Short Form
TLI: Tucker-Lewis Index
UK: United Kingdom